\documentclass{article}
\usepackage[dvips]{epsfig}
\usepackage[flushleft]{threeparttable}
\usepackage[english]{babel}
\usepackage[margin=2cm]{geometry}

\usepackage{amsmath}
\usepackage{graphicx}
\usepackage[colorinlistoftodos]{todonotes}
\usepackage[colorlinks=true, allcolors=blue]{hyperref}

\newcommand\fd{\hbox{$.\!\!^{\reset@font\romn d}$}}
\newcommand\fh{\hbox{$.\!\!^{\reset@font\romn h}$}}
\newcommand\fm{\hbox{$.\!\!^{\reset@font\romn m}$}}
\newcommand\fs{\hbox{$.\!\!^{\reset@font\romn s}$}}

\newcommand\farcs{\hbox{$.\!\!^{^{\prime\prime}}$}}
\newcommand\fp{\hbox{$.\!\!^{\reset@font\reset@font\scriptscriptstyle\romn p}$}}

\title{Observational Constraints on the Feeding of Supermassive Black Holes}
\author{Thaisa Storchi-Bergmann \& Allan Schnorr-M\"uller}

\begin{document}
\maketitle

\begin{abstract}

Supermassive Black Holes grow at the center of galaxies in consonance with them. In this review we discuss the mass feeding mechanisms that lead to this growth in Active Galactic Nuclei (AGN), focusing on constraints derived from observations of their environment, from extragalactic down to galactic and nuclear scales. At high AGN luminosities, galaxy mergers and interactions play an important role in AGN triggering and feeding. However, gas chaotic cold accretion in galaxy clusters can trigger radiatively inefficient AGNs in brightest cluster galaxies (BCGs). At lower luminosities, minor mergers feed AGN in early-type, gas-starving galaxies, while secular processes dominate in later-type, gas-rich galaxies. While bars do not appear to directly feed AGNs, AGN flickering leads to the dissociation between small and large scales, hence affecting the interpretation of cause and effect. At $\sim$100 pc scales, recent observations have revealed compact disks and inflows along nuclear gaseous spirals and bars, while chaotic cold accretion continues to feed BCGs at these scales. Estimated mass inflow rates -- of 0.01 to a few M$_\odot$\,yr$^{-1}$ -- are in many cases a thousand times higher than the mass accretion rate to the supermassive black hole. As a result, 10$^6$--10$^9$ M$_\odot$ gas reservoirs can be built on 10$^{7-8}$\,yr, that in turn may lead to the formation of new stars and/or be ejected via the onset of AGN feedback.

\end{abstract}



\section{Introduction}

The discovery of numerous correlations between the masses of supermassive black holes (SMBH) in the nuclei of galaxies and their host galaxies properties, such as the bulge velocity dispersion, mass and luminosity \cite{ferrarese00,gebhardt00,mcLure02,marconi03,haring04}, as well as the similar evolution of the star-formation and SMBH growth rates over cosmic time \cite{madau14}, has led to the conclusion that the SMBH and its host galaxy 
co-evolve \cite{kormendy13,heckman14}. Understanding how this co-evolution occurs is now a mAstron. J.or subject of research in extragalactic astrophysics. 

Feedback originated in the surroundings of SMBHs, as observed in Active Galactic Nuclei (AGN) has often been invoked as the process leading to the co-evolution of SMBHs and galaxies, as well as for regulating the galaxy growth to avoid producing overmassive galaxies \cite{fabian12, mcnamara12}. Feedback from the AGN, in the form of radiation, jets and accretion disk winds \cite{greene11,tombesi13,zubovas14,harrison18} may indeed heat and remove gas from the nuclear region of the host galaxy, suppressing star formation and regulating the growth of the galaxy, impacting also in the surrounding intergalactic medium \cite{gaspari12}.

The less studied feeding processes of SMBHs precede the feedback processes, and play a fundamental role not only on the SMBH evolution, but also in the evolution of its host galaxy. The discussion of these feeding processes and their observational constraints are the subject of the present review. 



The onset of nuclear activity depends on the gas supply to feed the AGN, that varies largely with its luminosity: luminous quasars (L$_{AGN}\ge10^{46}$\,erg\,s$^{-1}$) accrete tens of solar masses per year, while local Seyferts (L$_{AGN}\le10^{44}$\,erg\,s$^{-1}$) accrete only $\approx\,10^{-3}$\,M$_\odot$\,yr$^{-1}$. Considering a typical activity cycle of $10^{7-8}$\,yrs, while in low luminosity AGNs the gas needed to feed them can be supplied by the mere mass loss of evolving stars in the bulge \cite{ho08}, luminous AGN require large amounts of gas to be transported to the vicinity of the SMBH in a short period of time.  

Violent disturbances caused by major mergers are known to drive gas to the center of galaxies and fueling starbursts. Part of this gas could also fuel a SMBH in the center. Major mergers are thus an attractive option as a fueling mechanism of luminous quasars \cite{croton06,hopkins06}. In the case of lower luminosity AGNs, the fueling seems to occur via secular processes, internal to galaxies, which act on timescales longer than the dynamical time \cite{marleau13,reines15,simmons17}. The mapping of such inflows is now becoming available via Integral Field Spectroscopy (IFS) of nearby galaxies, in the optical and near-infrared wavebands, as well in the sub-millimetric spectral region via observations with the Atacama Large Millimetric Array (ALMA).

Although we include some discussion of the theory of gas inflows, this review will mainly focus on the observation of resolved gas inflows to AGN hosts from extragalactic through kiloparsec and down to hundred pc scales.  We will not discuss here unresolved studies of the inner parsec, where the ultimate inflow to the SMBH occurs via the torus, broad line region (BLR) and accretion disk. Recent studies have shown that the BLR frequently contains a flattened component in rotation that seems to be the outer parts of the accretion disk, thus suggesting the presence of inflow in these structures at sub-parsec scales \cite{elitzur14,pancoast14,SB17}.

\section{Inflows on extragalactic scales - the role of the environment}

\subsection{Major mergers}

Theoretical studies suggest that major mergers (with mass ratios of at most 4:1) are the dominant processes leading to SMBH growth at high masses -- M$_{SMBH}$\,$\ge$\,10$^8$\,M$_\odot$ \cite{menci14,hopkins14,gatti14}. At high redshifts ($z$\,$>$\,2), major mergers have also been proposed as fueling mechanisms of the fastest-growing SMBHs \cite{treister12}, although some studies have shown that not all high redshift quasars have companions \cite{fan17,trakhtenbrot17}, which suggests that secular mechanisms also play a role in feeding these objects in the early universe \cite{schawinski12}. This is further discussed bellow in Sec. \,\ref{secular}. 

A major merger is a phenomenon which can destabilize large quantities of gas, driving massive inflows towards the nuclear region of galaxies and triggering bursts of star formation. Many studies have found that the most luminous AGN (L$_{AGN}$\,$\ge$\,10$^{46}$\,erg\,s$^{−1}$) are preferentially hosted by galaxy mergers \cite{urrutia08,glikman12,glikman15,treister12,fan16}, in accordance with theoretical predictions. Major mergers have also been argued to trigger radio galaxies, such as B2 0648+27, for which observations in neutral gas (HI) \cite{emonts06a} have been used to date the merger, whose age was then compared with that of the stellar population and radio source. It was concluded that there was a delay of $\approx$ 1 Gyr between the merger and a starburst of age 0.3 Gyr, with a further delay in the onset of nuclear activity, suggesting that this galaxy corresponds to a phase in a presumed evolutionary sequence between Ultra-Luminous Infrared Galaxies (ULIRGs) and early-type galaxies.

Signatures of inflows have been observed in ULIRGs such as NGC\,4188, in this case with the Herschell Space Observatory \cite{gonzalez12} revealing redshifted molecular absorption lines interpreted as due to mass inflow to the nucleus at a rate of up to $\approx$\,12\,M$_\odot$\,yr$^{-1}$. More recently, Herschell observations  of the interacting system Arp\,299A  \cite{falstad17}, also revealed inflows to the nucleus that seems to harbor an AGN and/or a dense nuclear starburst. As these cases -- in which inflows are observed in mergers -- seem to be more the exception than the rule, it may be that the inflowing phase (that should precede the outflow phase) is shorter than that of the outflow, or it is harder to observe.

\begin{figure}
\center
\includegraphics[scale=0.8]{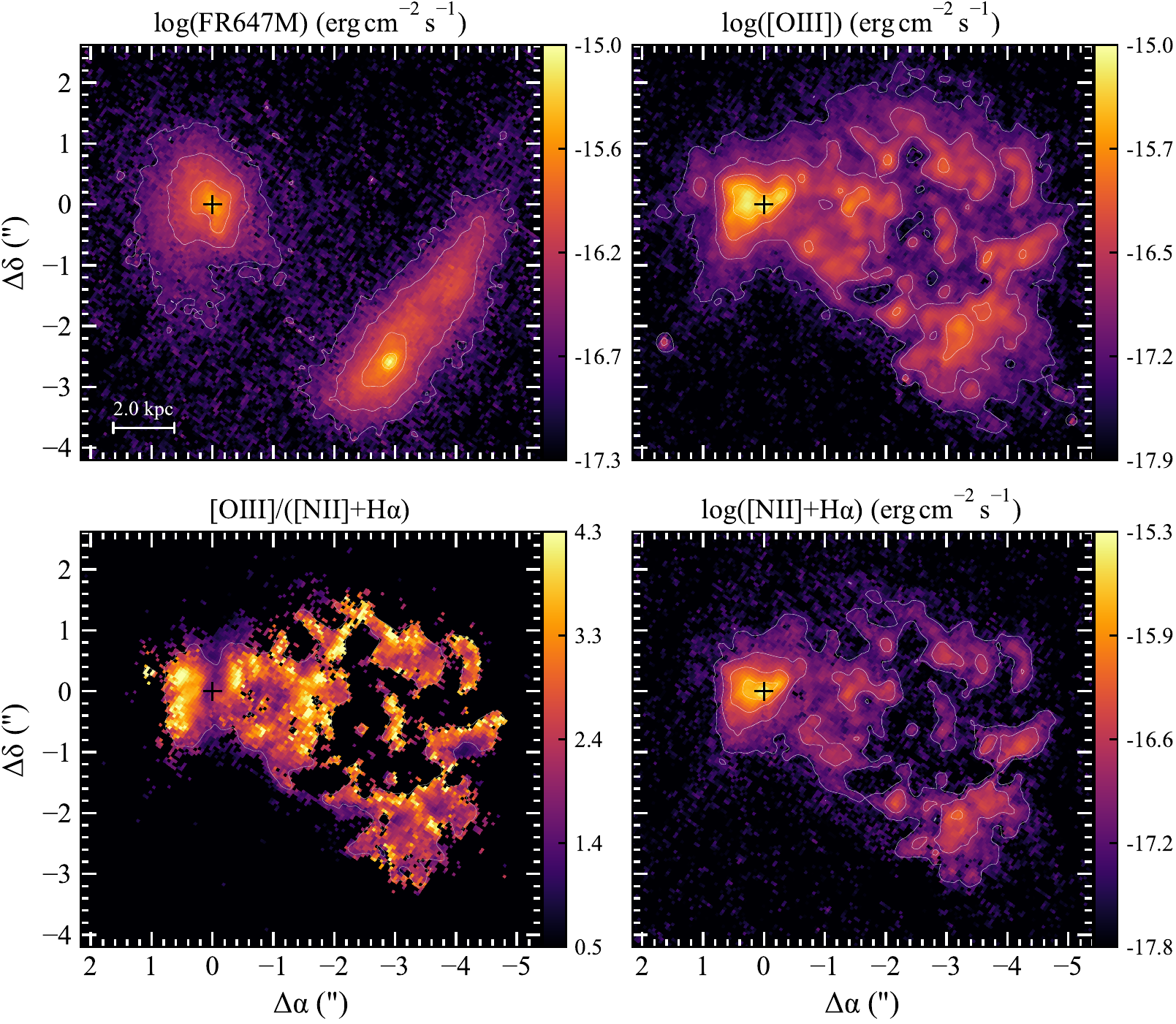}
\caption{Interaction between two galaxies triggering the nuclear activity in the QSO 2 SDSS-J084135.04+010156.3 as seen in HST images. Top left: continuum image; top right: [OIII]$\lambda$5007 narrow-band image; bottom right: H$\alpha$+[NII] narrow-band image; bottom left: excitation map obtained from the ratio between the two. Color bars give the fluxes in logarithm, in erg s$^{-1}$ units, except for the bottom left panel where the line ratio values are shown in linear scale. Contours in the two top panels and in the bottom right panel range from 3 times the root-mean-square standard deviation of the sky value in the vicinity of the galaxy to the maximum flux value. In the bottom left panel, the contour separates the regions of highest excitation (inside the contour), corresponding to [OIII]/([NII]+H$\alpha$)=1.3. North is up and East to the left in all panels. Credit: \cite{SB18}.}
\label{enlr}
\end{figure}

In a recent study \cite{SB18}, Hubble Space Telescope (HST) narrow-band images have been used to investigate the shape, excitation and extent of the narrow-line region (NLR) in nine luminous type 2 quasi-stellar objects. The sample was selected solely on the basis of its high AGN luminosity (L$_{AGN} \ge 10^{46}$\,erg\,s$^{−1}$) and it was found that most of the sources show signatures of interactions or mergers, also supporting the prevalence of mergers at high AGN luminosities. Figure\,\ref{enlr} shows the interaction of two galaxies potentially having led to the triggering of AGN activity at the centre of one of them.

At lower luminosities (L$_{AGN}\le10^{44}$\,erg\,s$^{-1}$), a number of studies have found a higher incidence of galaxies with signatures of interactions in AGN hosts as compared to control samples \cite{koss10} and in particular in close galaxy pairs \cite{silverman11,ellison11,ellison13,satyapal14} suggesting that kinematic pairs are conducive environments for black hole growth. 

Other studies have nevertheless found no increase in the incidence of major mergers by comparing the environment of AGN hosts to those of control samples of non-AGN \cite{boehm12,gabor09,grogin05,cisternas11,kocevski12, villforth14, karouzos14a,mechtley16,villforth17}. One effect that may cause conflicting results is the distinct timescale for the mergers and their observable signatures and those of the nuclear activity episodes \cite{hickox14}, as the SMBH accretion is not constant but may undergo several peaks during an interaction \cite{goulding18}.

Selection effects due to the use of different wavebands \cite{padovani17} (e.g. selection of samples via X-ray, optical, IR luminosity, radio) as well as different analysis techniques may also contribute to conflicting results. Recent studies and surveys show that most AGN in the Universe are actually hidden by gas and dust \cite{hick18}, what can bias also its detection.

A recent study has found that AGNs are five times more likely to be obscured when hosted by a merging system, which could lead to a significant fraction of merging AGNs being undetected at shorter wavelengths \cite{weston16a}. This is supported by another recent study comparing a sample of AGNs detected only in infrared wavelenghts to a sample of X-ray detected AGN \cite{donley17}, that  found that IR-only AGNs tend to be luminous, highly obscured and are more likely than X-ray detected AGN to be merging systems or to present irregular and assymetric morphologies. A similar result was recently obtained for mid-infrared selected AGN from the WISE all-sky survey and  high resolution images of the Hyper Suprime-Cam (HSC) survey of the SUBARU telescope \cite{goulding18}: applying a novel machine-learning technique, it was shown that galaxies undergoing mergers are more likely to contain luminous obscured AGN than non-interacting galaxies and that the most luminous AGN usually reside in merging systems. 

Finaly, it may be argued that, as mergers and interactions are known to boost star formation, and luminous AGN seems to correlate with recent circumnuclear star formation (within the inner few 100 pc), the relation between mergers and AGN could be a secondary one, resulting from the relation between mergers and circumnuclear star formation \cite{heckman14}.


\subsection{Minor mergers}
\label{minor}

Minor mergers -- adopted here as corresponding to galaxy mass ratios larger than 4, in which a massive galaxy captures a dwarf, gas-rich galaxy, can supply gas to an otherwise gas-starved galaxy. A recent theoretical study \cite{neistein14} using a semi-analytic model, has proposed that such minor merger events are important triggers of low to intermediate-luminosity AGN ($L\le10^{44}$\,erg\,s$^{-1}$). Some observations support this, at least for early-type galaxies \cite{sl07}: comparison of HST F606W images of the inner kiloparsec of local low-luminosity AGNs with those of a control sample has shown that all active early-type galaxies present dust structures in the inner kiloparsec, indicating the presence of cold gas, while these structures are present in only $\sim$\,25\% of the control sample. As the hosts of the AGN and controls are carefully matched, meaning that they can be considered ``the same galaxy seen in different moments", this result points to an external origin for the gas. The hypothesis of minor mergers as triggers of the AGN was then supported by a follow up study \cite{martini13} in which the corresponding dust masses of $\sim$ 10$^6$\,M$_\odot$ were obtained, consistent with those expected for gas-rich dwarf galaxies.


There are also detailed studies of radio galaxies in which the origin of the activity can be clearly attributed to a minor merger, the classical example being of Centarus\,A, which is estimated to have suffered such a merger a few $10^8$ yr ago \cite{struve10,morganti10}. Other examples include Arp\,102B, Pictor\,A and 3C\,33 \cite{couto13,couto16,couto17}. In the case of Pictor A, the inferred low metallicity of the gas further supports this conclusion.

Processes that are expected to occur as a result of the capture of minor galaxies have also been witnessed in a few studies of the internal kinematics of individual early-type galaxies. Examples include an integral field spectroscopic study using the instrument NIFS at Gemini that has found both ionized and warm molecular gas counter-rotating relative to the stars in the Seyfert 2 galaxy NGC\,5929 \cite{riffel15} and a similar study  using SINFONI at VLT that found a counter-rotating core both in stars and molecular gas in the Seyfert 1 galaxy MCG-6-30-15 \cite{raimundo17}.

A minor merger fueling an AGN was recently ``caught in the act'' in the vicinity of the nucleus of the Seyfert\,1 galaxy Mrk\,509 \cite{fischer15}: a peculiar HST image showed what seemed to be of a little galaxy falling into Mrk\,509. Observations of its kinematics, using the Gemini instrument NIFS then confirmed this scenario.

\subsection{Chaotic cold accretion}
\label{cool_accretion}

Besides major and minor mergers, accretion of cold gas streamers (cooling flows) from the hot intergalactic medium in galaxy clusters can also feed AGN. This seems to be the fuelling mode of the so-called mechanical-mode (or radio-mode) AGNs observed in radio galaxies (as opposed to the more luminous radiative-mode AGN). These objects are associated to the most massive SMBHs in the near universe, hosted by classical bulges, usually in the brightest central galaxies (BCGs) of rich galaxy clusters.

Recent models and simulations support that the gas reservoir feeding these SMBHs in the center of galaxy clusters consists of cold gas  \cite{pizzolato05,wada09,sharma12,gaspari14,li14,voit15a,voit15b}. In a turbulent halo of a galaxy cluster, as the hot plasma cools, warm filaments and cold clouds condense and rain towards the inner region being then chaotically accreted -- in a process known as Chaotic Cold Accretion (CCA \cite{gaspari13}). CCA has been defined as a rain of cold clouds condensing out of the quenched cooling flow and then recurrently funneled towards the SMBH via inelastic collisions \cite{gaspari15,gaspari17, gaspari18}. Recent ALMA observations show that molecular gas is a common presence in bright group-centered galaxies that seem to be the end product of the hot gas cooling process in their groups comprising unbound giant molecular associations drifting in the turbulent field, consistent with numerical predictions of the CCA process \cite{pasquale18}.


In further support to the above scenario, filaments of H$\alpha$ emitting gas have  been observed in a number of central cluster galaxies, the classical example being the radio galaxy NGC\,1275 at the center of the Perseus galaxy cluster \cite
{fabian12}. In a recent Gemini GMOS IFU study, similar inflows traced by H$\alpha$ filaments falling into the LINER nucleus of the elliptical galaxy NGC\,5044 have been reported \cite{diniz17}. ALMA observations of this galaxy \cite{david14} have revealed a number of molecular structures, concluded to be giant molecular cloud associations, including a redshifted absorption indicating infalling gas at 250\,km\,s$^{-1}$. In addition, cold (and warm -- at $\sim$2000\,K) molecular gas have been observed around the nucleus of other radio galaxies \cite{burillo07,labiano13,labiano14,tremblay16,russell16}. Signatures of cold H\,I disks have also been observed in the nuclear spectra of such galaxies  \cite{morganti09,maccagni14}, including the observation of redshifted H\,I absorption interpreted as due to cold clouds infalling towards the nucleus. 

Intergalactic gas has also been argued to be the fuel of radio-quiet early-type AGN. A study of the environmental dependence of AGN hosted by S0 galaxies in galaxy clusters \cite{davies17} showed that their fraction decreases in large clusters relative to smaller galaxy groups, despite the increase of the fraction of S0 galaxies in these richer environments. This result has been interpreted as an evidence of an external origin for the gas supply in S0 galaxies: while the intra-group gas would be relatively cold in smaller groups and thus easily accreted, its higher temperature in denser environments would decrease its accretion efficiency. For the later-type spiral galaxies, no environment dependecy was observed, supporting the prevalence of secular processes for these galaxies.

\subsection{Further accretion of intergalactic gas} 

Recently,  yet another environmental mechanism resulting in AGN feeding has been proposed \cite{poggianti17}:  6 out of a sample of 7 ``jellyfish" galaxies in a galaxy cluster have been found to show signatures of radiative-mode AGN at their nuclei.  Such galaxies are called ``jellyfish" because they show long ``tentacles"  due to strong ram pressure stripping within galaxy clusters.  This result has been interpreted as due to the interaction with the intra-cluster medium causing the gas in the AGN host to lose angular momentum and flow inwards. Although such processes have been predicted by simulations \cite{marshall18}, many studies report that in galaxy clusters, most radiative-mode AGNs are located at large radii \cite{ruderman05,haines12,pimbblet13,pentericci13,ehlert13,mahajan12}, with the AGN fraction decreasing at smaller radii \cite{gilmour07,gavazzi11,pimbblet12}. Consequently, while ram pressure stripping might trigger nuclear activity in infalling galaxies, it inhibits radiative-mode nuclear activity in galaxies that are near the cluster core \cite{gordon18}.

\section{Inflows on galactic scales: secular processes}
\label{secular}

Theoretical studies suggest that galaxy disk instabilities and associated inflows -- secular processes occurring on evolutionary timescales, larger than the dynamical timescale -- are the dominant  processes leading to SMBH growth at low masses -- M$_{SMBH}$\,$\le$\,10$^7$\,M$_\odot$ \cite{hopkins14} in radiative-mode AGN. Secular processes lead to the formation of pseudo-bulges \cite{kormendy13}, and these have indeed been observed in many late-type active galaxies, notably in the particular case of Narrow-Line Seyfert\,1 galaxies \cite{xivry11}.

Observational support for the prevalence of secular processes in the feeding of AGN hosted by late-type galaxies include recent studies of the cold molecular gas content in luminous nearby AGN hosts. One of these studies -- the Luminous Local AGN with Matched Analogs (LLAMA) survey \cite{davies15,rosario18} shows that, although the AGN hosts display elevated levels of central star formation compared to inactive galaxies of similar stellar mass and Hubble type, there seems to be no difference in gas fractions and central star-formation efficiencies between active and inactive  galaxies. These results seem to imply that in late-type galaxies, secular processes are always in action, and AGN would be triggered in association with star formation: whatever triggers star formation in the circumnuclear region, triggers also the nuclear activity.

\subsection{Distant galaxies ($z>1$)}

Recent observations suggest that secular processes may also feed AGN in the distant Universe (z\,$>$\,1), as supported by the high incidence of AGN in massive, star-forming disk galaxies at these redshifts \cite{genzel14}. 

Other studies have pointed out the frequent presence of giant star-forming clumps (10$^8$-10$^9$\,M$_\odot$, 100-1000\,pc) in distant disk galaxies \cite{cowie96,elmegreen04,elmegreen05,schreiber06,ravindranath06,genzel08,guo12}, that are supposed to be formed via fragmentation of their gas rich disks due to gravitational instabilities \cite{shapiro08,bournaud08,schreiber09,genzel11,mandelker14}. These clumps are the sites of most star formation in their host galaxies, and, if long lived, will sink to the nucleus due to dynamic friction and potentially feed an AGN \cite{bournaud11}. Alternatively, gravitational torques induced by the clumps can drive the gas between them inwards \cite{dekel13a}. 

Observations point to clumps being short-lived (100-200\,Myr, \cite{wuyts12}), meaning they can only reach the center if their migration is efficient, and suggest only the densest clumps survive the dynamical interactions with the disk and reach the center \cite{cava18}. A short lifetime for star forming clumps is supported by most simulations \cite{hopkins12,hopkins13,forbes14,oklopcic17}, although the most massive and dense ones may  survive for hundreds of Myr \cite{bournaud14} and migrate to the center of the disk \cite{mandelker17}.

The observational picture is still inconclusive and in need of more studies: while an increase in the AGN fraction hosted by galaxies with clumpy disks compared to smooth disks has been found at redshift z\,$\sim$0.7 \cite{bournaud12}, no such increase has been found in a z\,$\sim$\,2 sample \cite{trump14}.

\subsection{Galactic scale bars}

Galactic  bars are a common feature in galaxy disks in the local Universe. They have long been proposed as a mechanism for sending gas to the nucleus and leading to the triggering of nuclear activity \cite{schlosman89}, as they can efficiently transfer angular momentum within the disc, allowing gas to move inwards \cite{athanassoula92,regan95,regan99,sakamoto99,sheth05}. Recent models \cite{li15} show that, due to the torque of the bar, the gas in the galaxy develops shocks traced by dust-lanes at the leading side of the bar and flows in, frequently forming a ring at a few 100 pc from the nucleus, approximately at the location of the Inner Lindblad Ressonance  \cite{knapen06}. 
The gas that flows down the bar seems to stay ``trapped'' in the ring. Another mechanism needs to be present to promove further inflow -- within the inner few 100 pc -- in order to feed an AGN. A few galaxies with nuclear rings show nuclear activity \cite{sb96b,sb96a,fathi06,hennig18}, that could be interpreted as due to triggering by this further inflow.

In order to evaluate the relation between nuclear activity and galactic-scale bars, many studies searched for an enhanced bar fraction in AGN hosts as compared to control samples of spiral galaxies. A large number of such studies concluded that there is no such enhancement in the near Universe \cite{ho97,mulchaey97,malkan98,hunt99,martini99,erwin02,lee12}. A smaller number of studies find an enhancement, but with a small significance, of $\le2.5\sigma$ \cite{knapen00,laine02,laurikainen04,alonso13}. A recent investigation looking for the incidence of bars in galaxies with intermediate redshifts (0.2\,$<\,z\,<$\,1)\cite{cheung15} did not find excess of bars in AGN hosts, concluding also that there is no evidence that galactic-scale bars fuel AGNs.

The relation between nuclear activity and bars has been further explored via a comparison between accretion rates in barred and non-barred AGN hosts \cite{cisternas13,galloway15}, that resulted in comparable values. A similar result was obtained in a study using Chandra X-ray archival data to derive an average X-ray luminosity for barred and non-barred galaxies through stacking analysis  \cite{goulding17a}, that did not show any systematic difference in the average X-ray luminosity between them. These studies imply that AGN are not fed -- at least directly -- via galactic scale bars.


\section{Inflows on 100 pc scales}

\begin{figure}
\center
\includegraphics[scale=0.5]{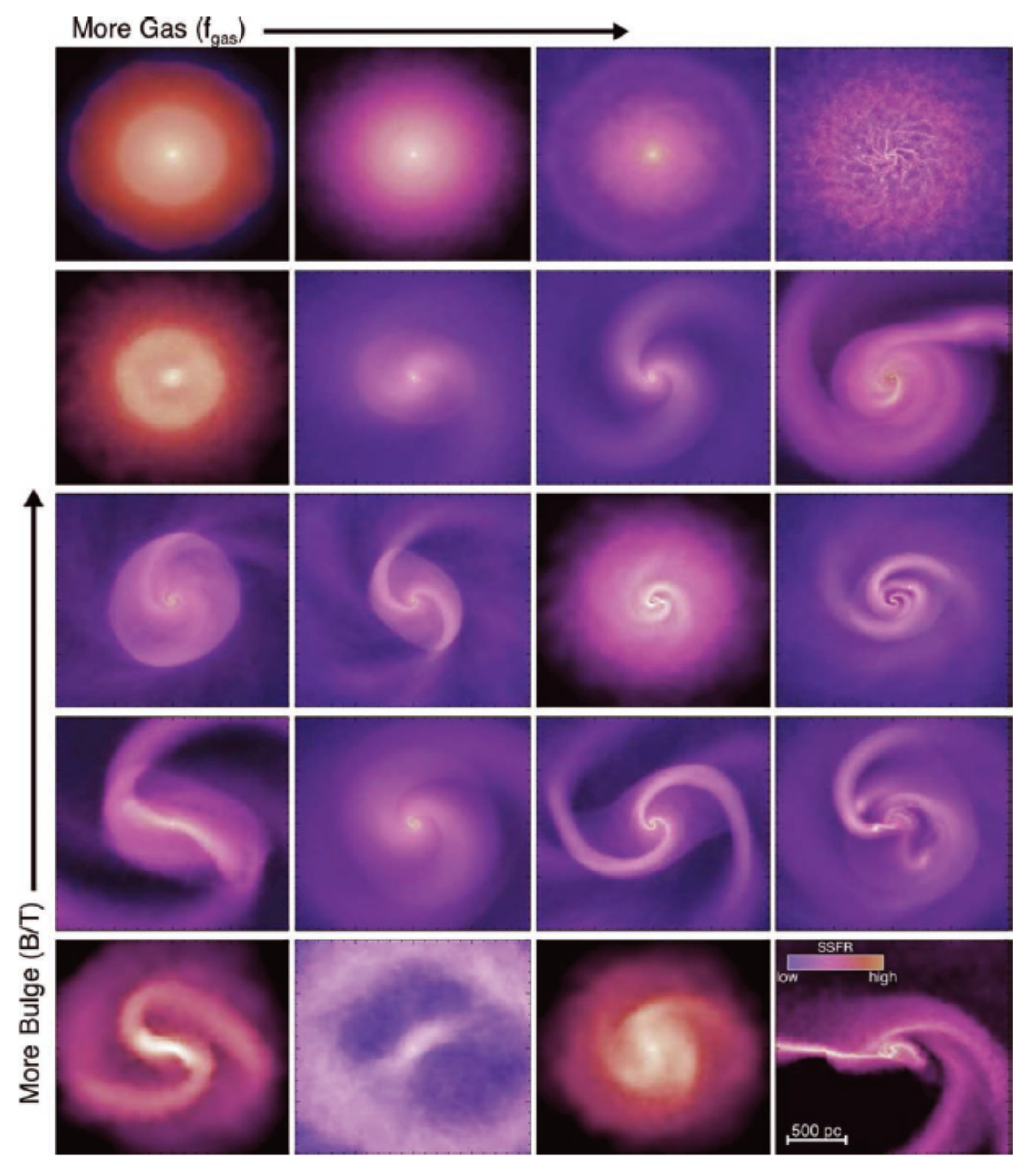}
\caption{ {\bf A model for inflows at high AGN luminosities ($\ge\,10^{46}$erg\,s$^{-1}$) on ~ few 100\,pc scales:} reproduction of Fig. 3 from Hopkins et al. (2010) \cite{hopkins10}, showing the gaseous structures developing as a consequence of galactic-scale torques bringing gas into the central kpc. Each panel shows projected gas density that is proportional to the local Specific Star Formation Rate (colour, from blue--low, through yellow--high; typical star formation rates are $\approx$\,10 to a few 10's M$_\odot$\,yr$^{-1}$ ) with bulge (B) to total (T) mass ratios decreasing from top (B/T$\ge$0.8) to bottom  (B/T$\le$0.1), and with gas fraction ($f_{gas}=M_{gas}/(M_{gas}+M_{stars,disk}$) increasing from left ($f_{gas}\le0.1$) to right ($f_{gas}\ge0.8$).}
\label{hopkins_10}
\end{figure}

\begin{figure}
\center
\includegraphics[scale=0.4, angle=-90]{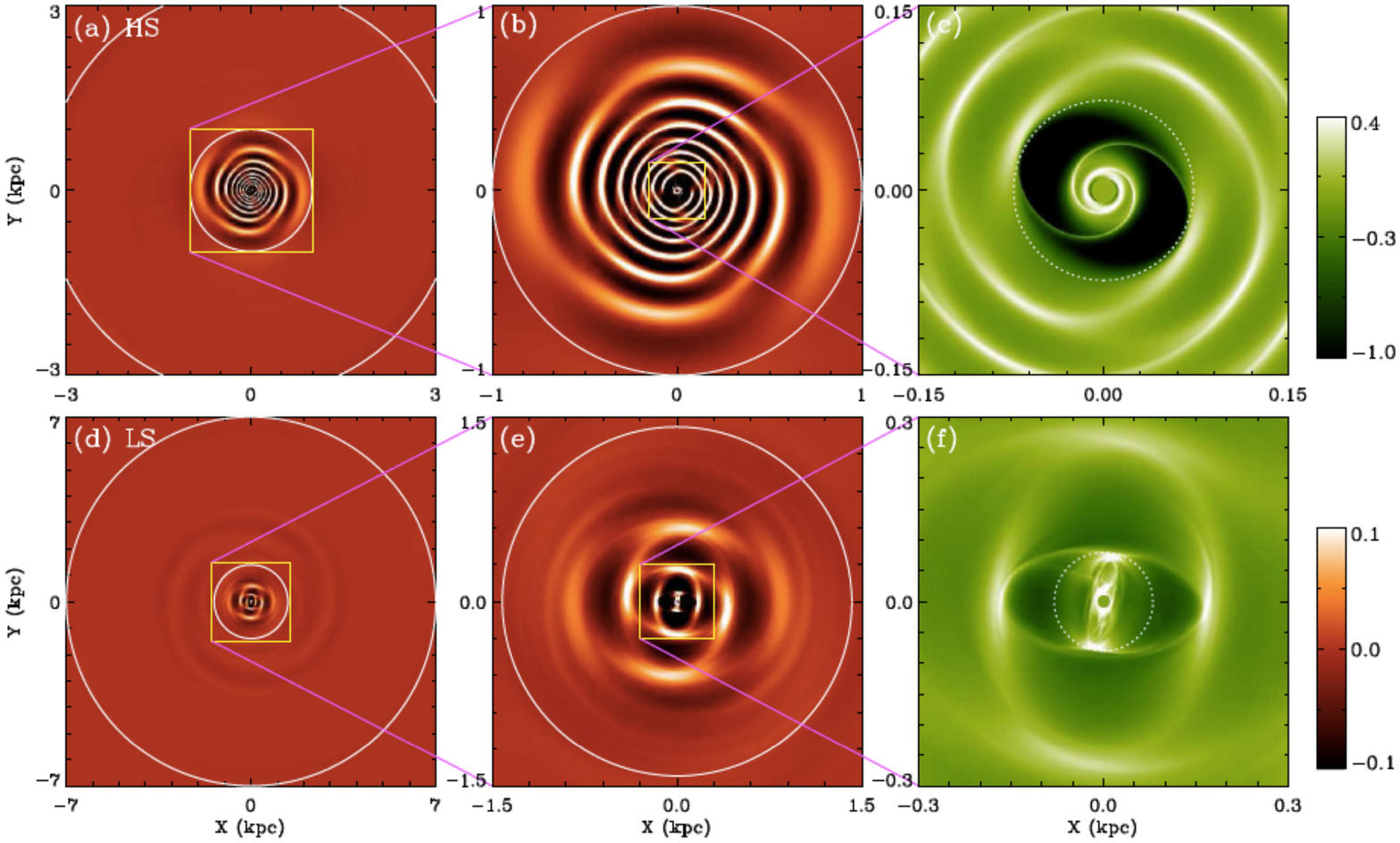}
\caption{ {\bf A model for inflows at low AGN luminosities ($\le\,10^{44}$erg\,s$^{-1}$):} reproduction of Fig. 4 from Kim \& Elmegreen (2017)   \cite{kim17}, showing the gas surface mass density distribution $\Sigma/\Sigma_0$, where $\Sigma_0$ is the initial surface density, for two typical cases: high shear (top panels) and low shear (bottom panels). In the right panels, the dotted circle at about 80\,pc marks the boundary of the region dominated by shocks. Color bars show the variation of log($\Sigma/\Sigma_0$), with the upper (lower) one corresponding to the right (left and middle) panels.}
\label{fig:kim}
\end{figure}

Models of gas inflows to feed SMBHs show that the relevant processes occur within the inner kiloparsec. Two of these models\cite{hopkins10,kim17} are discussed in Box 1, and illustrated in Figs.\,\ref{hopkins_10} and \ref{fig:kim}. We discuss bellow the observational signatures of inflows on 100\,pc scales, which can be compared to these models.

Previous theoretical studies \cite{fukuda98,maciejewski02,maciejewski04} have also developed hydrodynamic models of gas flows towards a central SMBH, using a bar-like perturbation on large scales to cause the initial inflow, and also show that, within the inner kpc,  gas can flow to the center in spiral shocks. These studies have also shown that a central SMBH can allow the spiral shocks to extend all the way to its immediate vicinity, and to generate gas inflows of up to 0.03\,M$_{\odot}$\,yr$^{−1}$, which is of the order of the accretion rates needed to power local low-luminosity AGN. 

A recent simulation of a Milky Way-like galaxy has followed the triggering of inward gas motion towards inner resonances via a large-scale bar and the connection to the central black hole via mini spirals on hundred of pc scales in a self-consistent manner \cite{emsellem15}.

The mechanism of chaotic cloud accretion -- CCA \cite{gaspari17} can also apply to the inner 100\,pc of galaxies, most notably to the massive early-type galaxies at the centers of galaxy clusters. The recurrent collisions and tidal forces between clouds, filaments, and in particular in  the central clumpy torus promote angular momentum cancellation, hence boosting accretion such that on sub-pc scales the clouds are channelled to the very centre via a funnel. The processes repeat in a fractal way and may reach down to 20 gravitational radii \cite{gaspari13}. Another interesting characteristic of CCA is also the possibility that the clumps also form stars.

On scales of tens of parsecs, down to the torus scales, recent models include the works of \cite{vollmer08} and \cite{vollmer13}, in which the authors propose how compact circumnuclear molecular gas disks form and evolve around galaxy centers, undergoing gravitational instabilities, the formation of clumps and of new stars \cite{kawakatu08,schartmann18}.

We now discuss observational studies showing the above mechanisms in action.

\subsection{Tracing the fueling mechanisms: imaging}

\begin{figure}
\center
\includegraphics[scale=0.3]{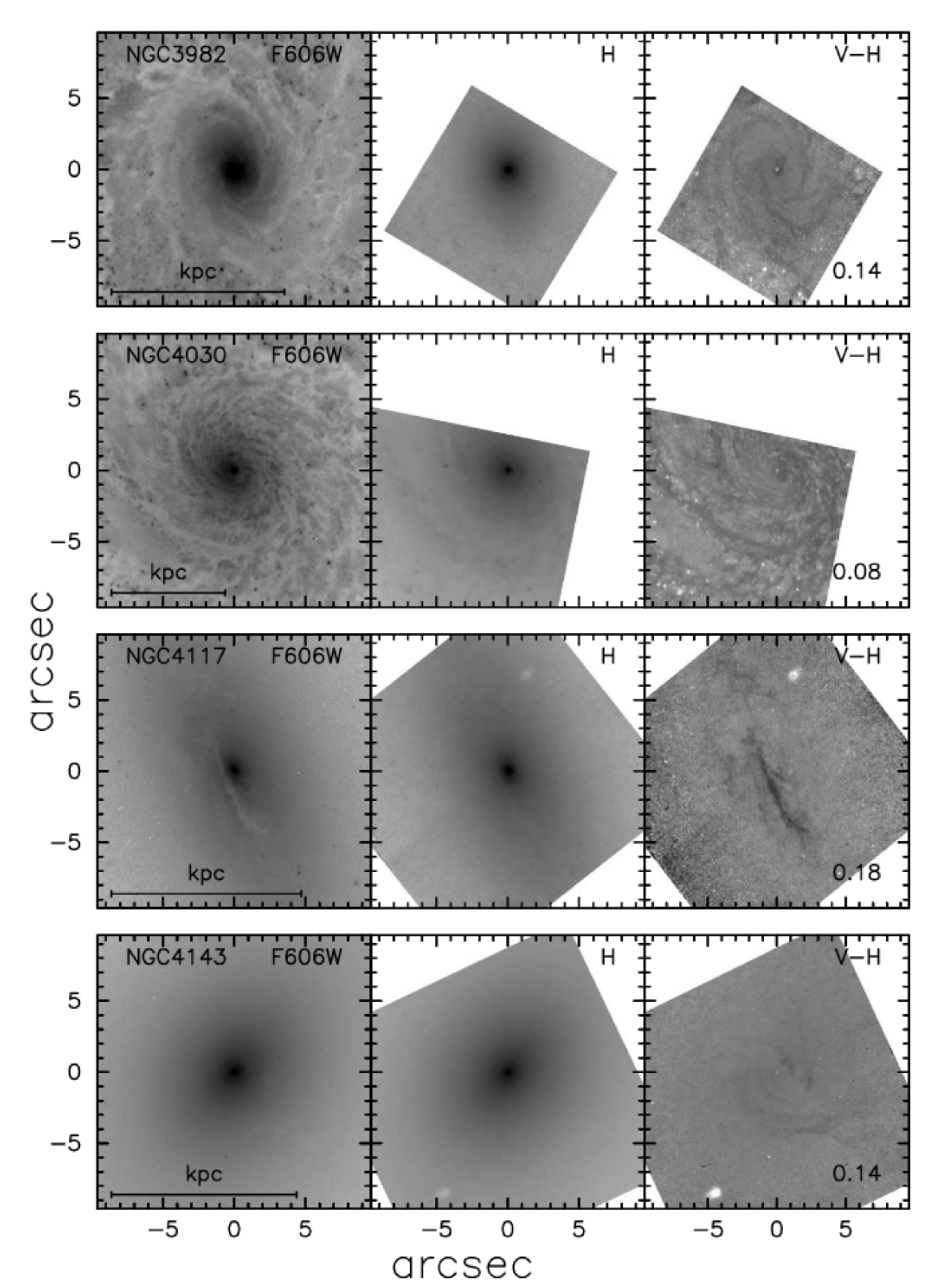}
\caption{{\bf The inner kiloparsec structure of a sample of active galaxies} as seen in HST F606W (V, left column) and NICMOS H band images (central column), as well as in a color map V-H (right column), where dark correspond to red color and bright to a blue color, showing the prevalence of dusty nuclear spirals within the inner kpc of AGN hosts.  Reproduction of part of Fig.\,1 of Martini et al. 2003 \cite{martini03}.}
\label{martini03}
\end{figure}

The superb image quality of the Hubble Space Telescope motivated several studies of the circumnuclear structure (inside the inner kpc) of nearby AGN hosts with the aim of identifying their feeding channels \cite{malkan98,regan99,martini99,martini01,pogge02, sl07}. Such studies show that dusty nuclear spirals are a common feature of the inner few hundred pcs of the majority of active galaxies, as illustrated in Fig.\,\ref{martini03} \cite{martini03}, suggesting that they play a major role in the fueling of AGN. Another common feature of the vicinity of nearby AGN are nuclear bars, present in $\approx$\,25\% of Seyfert galaxies \cite{regan99,martini01}. 

An imaging study of the inner kiloparsec of the LINER/Seyfert\,1 galaxy NGC\,1097 \cite{prieto05}, including a comparison of the nuclear spirals within its $\approx$\,700\,pc star-forming ring with models, show that  these spirals are the channels through which matter is being transferred to the nucleus to feed the AGN.  

The marked difference between early-type AGN hosts and control sample within the inner few 100 pc \cite{sl07}, discussed in Sec.\,\ref{minor},  in which all AGN hosts present nuclear dusty structures (in its majority nuclear spirals), but only $\approx$\,25\% of the control sample present them, also support the hypothesis that the nuclear structres/spirals are the feeding channels to the AGN, in agreement with models \cite{kim17}. In addition, the difference between AGN hosts and control sample implies that the presence of gas reservoirs on hundred pc scales is a necessary condition for the triggering of nuclear activity, and, in addition, that this gas has to be depleted in one activity cycle.

\subsection{Tracing the fueling mechanisms: kinematics}

Although images do indicate the presence of nuclear spirals in nearby AGN hosts, stronger constrains on inflows are given by the gas kinematics, which we discuss below.



\subsubsection{Cold molecular gas} 


Cold molecular gas is abundant in the inner few 100 pc of radiative-mode AGN \cite{burillo04,mazzalay13}. The mapping of its kinematics has been done in recent years by the NUGA (Molecular gas in NUclei of GAlaxies) group \cite{burillo12,burillo15}, using observations with the IRAM Plateau de Bureau Interferometer (PdBI) in the molecular emission lines $^{12}$CO(1-0) and $^{12}$CO(2-1). The spatial resolution at the galaxies range from a few tens to hundred parsecs. Combining these observations with images of the galaxies in the K band, the  NUGA group has calculated the torques acting on the gas. Negative torques -- implying inflows -- on hundred of pc scales have been found in a number of galaxies \cite{lindt08,hunt08,casasola08,burillo09,casasola11}, comprising 1/3 of a sample of 25 galaxies \cite{burillo12}, that is not restricted to AGN, but include a few of them. Although the investigated torques are important agents of gas inflow, the theoretical studies discussed above show that other effects should also be taken into account such as the development of shocks in nuclear spirals allowing for the gas to flow in, as well as the gravitational effect of the SMBH at the center.

Although not explicitly showing inflows, PdBI observations of dense HCN molecular gas in 4 Seyfert galaxies \cite{sani12,lin16} have revealed the presence of compact ($\sim$\,100\,pc) dense circumnuclear gas that is presumably the source of fuel of the AGN. Another nearby AGN showing a circumnuclear disc (within the inner 400\,pc) of cold gas is Centaurus A \cite{morganti10}, where filaments and streamers extending down to the inner 10\,pc have been recently observed with ALMA (Atacama Large Millimetric Array) \cite{espada17}. 

\begin{figure}
\center
\includegraphics[scale=0.35]{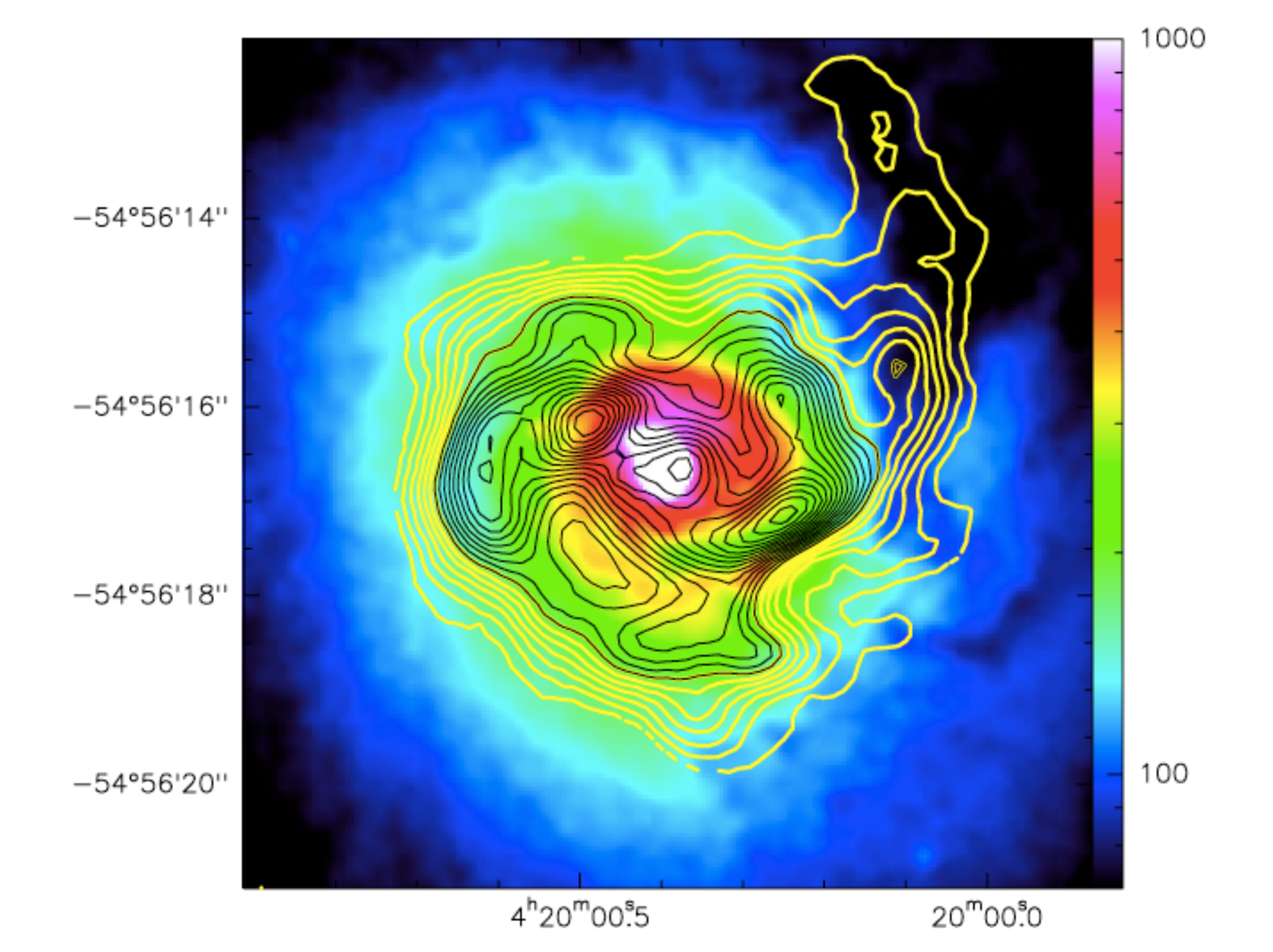}
\caption{{\bf Inflow in the Seyfert\,1 galaxy NGC\,1566 observed with ALMA:} Overlay of CO($3-2$) contours on the (F606W) HST image of NG\,1566 showing that the ALMA CO emission coincides with a dust spiral arm to the NW. CO contours are linear from 0.04 to 17.44 in steps of 0.87 Jy/beam\,km\,s$^{-1}$. Scale at the galaxy: $\approx$\,50\,pc/${^{\prime\prime}}$. Reproduction of Fig.\,4 of Combes et al. (2014)\cite{combes14}. }
\label{combes14}
\end{figure}

Other galaxies for which inflows have been found with ALMA include the LINER NGC\,1433 \cite{combes13} and the Seyfert\,1 NGC\,1566 \cite{combes14,slater18}. In NGC\,1566, molecular gas inflows between 300 and 50\,pc from the nucleus have been found in association with a nuclear spiral (see Fig.\,\ref{combes14}), as revealed by the inferred negative torques. In the more distant radio galaxy PKS\,B1718-649, signatures of inflow have been seen in ALMA observations of cold molecular gas and seem to be associated with previous inflows seen in H\,I and warm molecular gas \cite{maccagni18}. 

Episodes of chaotic cold accretion (CCA) have also been argued to be found at about 100\,pc from the nucleus of brightest cluster galaxies (BCG), an example being the BCG at the center of the cluster Abell 2597 \cite{tremblay16}. Absorption lines seem in ALMA observations towards the nucleus indicate inflows at speeds of $\approx$\,300\,km\,s$ ^{-1}$ that are corroborated by higher resolution warmer atomic gas data that imply the above small distance from the nucleus.

Inflows in atomic gas have also been found in optical spectra \cite{krug10}, observed as redshifted absorptions in NaI\,D ($\lambda 5890\AA$). Such absorptions have been found in about 1/3rd of a sample of 35 Seyfert galaxies selected for being  ``infrared-faint" (not LIRGs -- Luminous Infrared Galaxies), what could suggest that these galaxies have been caught in a phase in which the gas is still predominantly accreting into the central region.

Finally, the detection of inflows in molecular gas with ALMA has been recently reported also for the Milky Way nucleus \cite{hsieh17}  in the form of  dense molecular streamers in CS(J = 2 - 1) emission that seem to originate from ambient clouds at $\approx$\,20\,pc from the nucleus moving towards the central circumnuclear disk at 2\,pc. At the time of writing, many studies are being done with ALMA, in particular new high angular resolution studies that may reveal more molecular gas inflows towards the center of AGNs in the near future.

\subsubsection{Warm molecular gas}
\label{warm_gas}

Inflows have been reported also in warm molecular gas (T$\sim$2000K) via the K-band H$_2\lambda2.122\mu$m emission line. The  AGNIFS (AGN Integral Field Spectroscopy) group (https://sites.google.com/view/agnifs) has used the Gemini Near-Infrared Integral Field Spectrograph (NIFS) to map the stellar and warm molecular gas kinematics of the inner few 100 pc of a sample of nearby 29 active galaxies, at spatial resolution (agular resolution $\approx$0\farcs1) ranging from 10 to 100 pc and velocity resolution of $\approx$\,40\,km\,s$^{-1}$. Although this warm molecular gas is just the ``warm skin" of a colder gas reservoir, the observations reveal the molecular gas kinematics, that frequently present signatures of inflows towards the nucleus. 

One method used to look for inflows is the fit of the stellar velocity field by a rotation model, then use this model to fit the gaseous velocity field allowing for variation in the rotation amplitude that is usually higher for the gas (due to its lower velocity dispersion as it is usually more confined to the galaxy plane than the stars), and then subtract this model from the gasesous velocity field \cite{diniz14}. The cases of inflows are revealed when the residuals show mostly blueshifts in the far side of the galaxy and redshifts in the near side. If the gas is in the galaxy plane, this kinematics imply inflows towards the center.

Although deviations from circular rotation are expected if the orbits are not circular (e.g. elliptical), two systematic results of the studies above -- (1) the association of the residuals with nuclear spiral arms seen in structure maps; and (2) the fact that the residuals are systematically observed as blueshifts in the far side and redshifts in the near side -- lend further support to the interpretation of inflows. If the residuals were mostly due to elliptical orbits, one would expect similar frequency of blueshifts and redshifts, irrespectively to the side of the galaxy plane the residual is observed (near or far). Residuals attributed to inflows are also less symmetric than those expected, for example, for gas in elliptical orbits or moving in a bar potential.

\begin{figure}
\center
\includegraphics[scale=0.5]{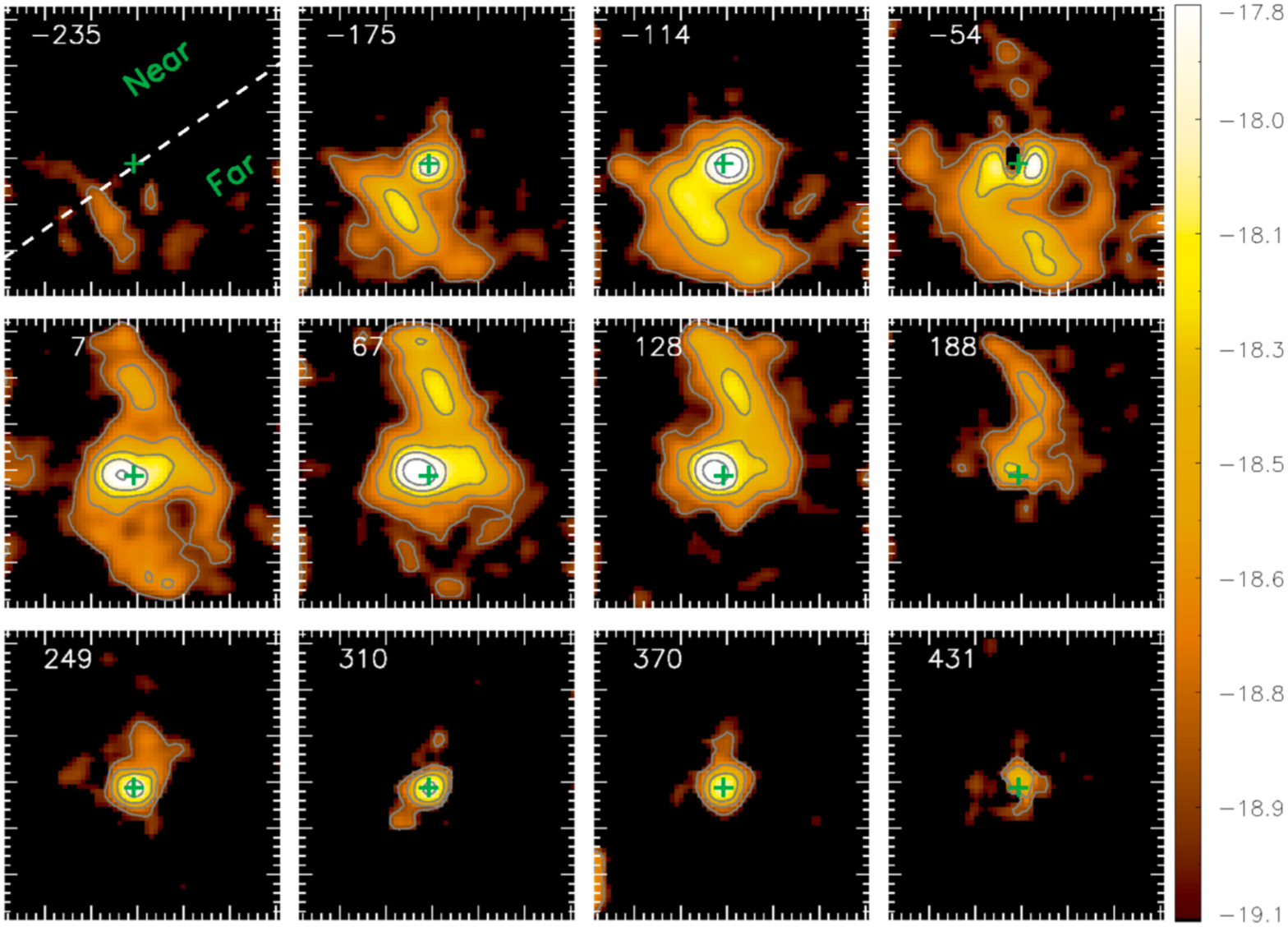}
\caption{{\bf Channel maps in the H$_2 \lambda 2.12\mu$m emission-line of the inner 600\,pc of the Seyfert\,1 galaxy Mrk\,79} for a velocity bin of 50\,km\,s$^{-1}$ and centered at the velocity shown in the top-left corner of each panel in km\.s$^{-1}$. Large tick marks are separated by 0\farcs5 (223\,pc at the galaxy), and the flux scale is logarithmic (log$_{10}$)  and units are ergs\,cm$^{-2}$\,s$^{-1}$. Contours are included (at arbitrary levels) only to highlight the distribution of the emitting gas at different flux levels. The cross marks the position of the nucleus, and the dashed line marks the kinematic major axis, with the near and far sides of the galaxy identified \cite{riffel13}.}
\label{mrk79}
\end{figure}

Signatures of inflows are also seen in channel maps on the H$_2\lambda2.122\mu$m emission line. One example is shown in Fig.\ref{mrk79} for the inner 600 pc of the Seyfert 1 galaxy Mrk\,79  \cite{riffel13}, in which such channel maps present mostly blueshifts in the far side of the galaxy and redshifts in the near side in a two-armed spiral structure. Again, if the gas is in the galaxy plane, this can be interpreted as due to inflows towards the center. This is distinct from a ``normal" rotation signature, in which the higher velocities are observed along the kinematic major axis and are symmetric relative to this axis. Other similar cases of gas inflow in nuclear spirals were mapped with NIFS around the nuclei of the active galaxies NGC\,4051 \cite{riffel08}, Mrk\,1066 \cite{riffel11} and NGC\,2110 \cite{diniz14}. 


Inflows in the H$_2\lambda2.122\mu$m line have also been found using IFS with the instrument SINFONI at the Very Large Telescope (VLT). This is the case of the inner few 100 pc of the Seyfert\,1/LINER galaxy NGC\,1097 \cite{davies09}, in which SINFONI observations, compared to results of hydrodynamic modelling led to the conclusion that inflows were occuring along nuclear spirals. Similar observations of NGC\,1068 \cite{mueller-sanchez09} have revealed molecular gas streamers towards the AGN at an estimated mass inflow rate of several M$_\odot$\,yr$^{-1}$. More recently, inflows along nuclear spirals were also found in the nearby Seyfert 2 galaxies NGC\,5643 and NGC\,7743, along a nuclear bar in the Seyfert 1 galaxy NGC\,3227 \cite{davies14}, and towards the nucleus of the radio galaxy PKS\,B1718-649 \cite{maccagni16}.

In some cases, spatially resolved inflows are not clear in the mapped kinematics, but compact rotating disks are seen instead. This has been seen in SINFONI observations of Cen A \cite{neumayer07} that revealed a rotating disk of radius $\approx$\,30\,pc. In NGC\,4151, a 50\,pc radius rotating disk seen at the inner end of the bar \cite{sb10} favor an origin from inflows along the bar, in agreement with models. Similar $\sim$\,30\,pc H$_2$ disks with estimated masses of 10$^7$\,M$_\odot$, have also been found in observations of a sample of nearby active galaxies via IFS with the instrument OSIRIS at the Keck telescope \cite{hicks09} and Gemini NIFS  \cite{riffel11,riffel11b}. The conclusion of these studies is that  the origin of such compact disks are massive inflows towards the nuclear region.

A comparison between the molecular gas properties of the inner 500\,pc of 5 active galaxies and a control sample using SINFONI data \cite{hicks13} revealed lower stellar velocity dispersions,  elevated H$_2\lambda2.122\mu$m luminosity and more centrally concentrated H$_2$ surface brightness profiles in the AGN hosts than in the control sample within the inner 200\,pc. These differences were interpreted as evidence for Seyfert galaxies having a dynamically colder nuclear structure than the bulge, composed of a significant gas reservoir and a relatively young stellar population recently formed from this gas. 

\subsubsection{Ionized gas}
\label{ionized}

In the optical, the AGNIFS group has been using the GMOS--IFU to map inflows using H recombination and [N\,{\sc ii}] lines, applying the same method described above for the warm molecular gas: the fit and subtraction of a rotation model.

One such case is that of the galaxy NGC\,7213 \cite{sm14}, hosting a LINER/Seyfert 1 nucleus, that shows velocity residuals that imply inflows in nuclear spirals seen in a structure map of the optical HST F606W image of the inner few hundred pcs of the galaxy. Other similar cases include the Seyfert galaxies NGC\,1667 \cite{sm17a}, NGC\,1358 \cite{sm17b}, and LINERs NGC\,1097 \cite{fathi06}, M81 \cite{sm11} and NGC\,6951 \cite{sb07}. In the case of the Seyfert 2 galaxy NGC\,3081 \cite{sm16}, gas inflows have been observed along a nuclear bar.

\begin{figure}
\abovecaptionskip-14ex
\belowcaptionskip-2ex
\includegraphics[scale=0.6, trim= 0 8 0 5cm]{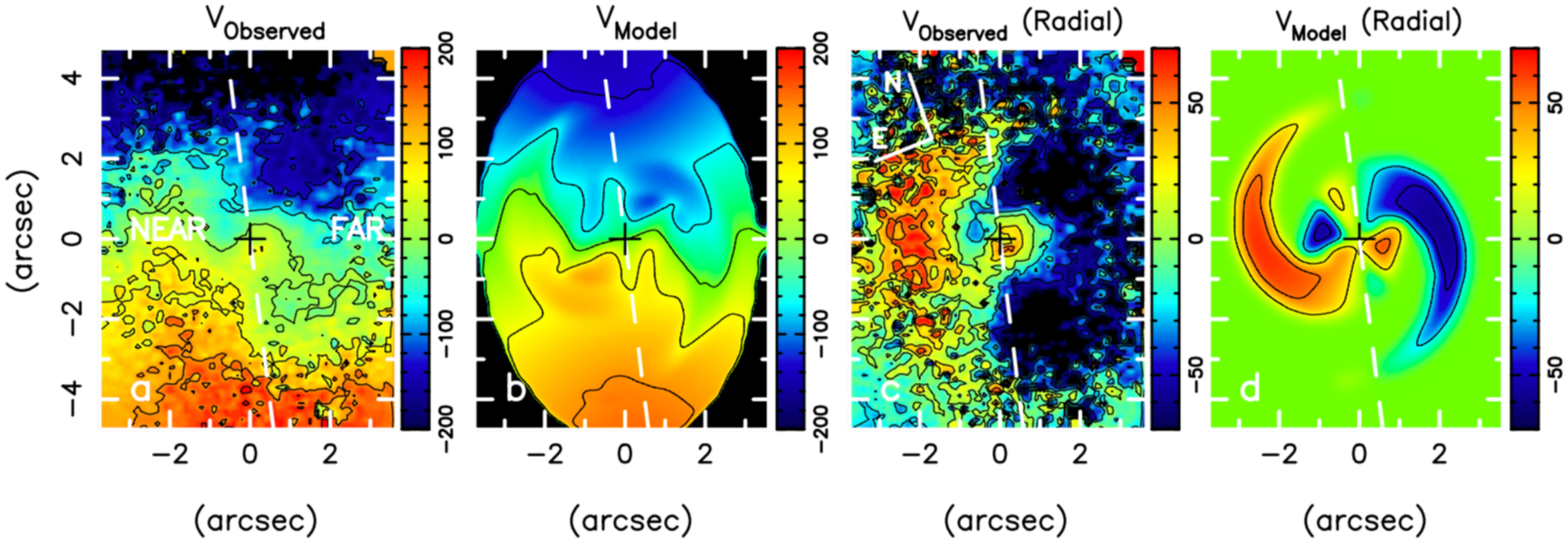}
\caption{{\bf Measured versus toy model kinematics of the inner kiloparsec of NGC\,1667 \cite{sm17a}}, showing: (a) measured gas velocity field, showing velocities according to the color bar in units of km\,s$^{-1}$; spatial scale is $\approx$\,300\,pc/${^{\prime\prime}}$; (b) toy model velocity field combining inflow in spiral shocks with rotation (plus a compact outflow at the nucleus); contours in panels (a) and (b)  are equally spaced and range from $-$200 to 200\,km\,s$ ^{-1}$; (c) measured gas velocities after subtraction of the rotation component showing only the residual inflows (plus compact outflow), that can be compared with the (d) inflow (plus compact outflow) component of the toy model; contours in panels (c) and (d) are equally spaced and range from $-$50 to 50\,km\,s$ ^{-1}$. The dashed line shows the line of nodes of the rotation component, and the near and far sides are identified in panel (a).}
\label{shape_ngc1667}
\end{figure}

In the recent work mentioned above in which inflows have been found towards the Seyfert 2 nucleus of the galaxy NGC\,1667 \cite{sm17a}, a phenomenological `toy model" using the software ``Shape" \cite{steffen11} was proposed for the gas kinematics, in which rotation in the disk was combined with radial inflows at the locations of the spiral arms, under the physical motivation that the dusty spirals are signatures of shocks, allowing the gas to lose angular momentum and move inwards \cite{kim17}.  A comparison between the observed and model velocity fields is presented in Fig.\,\ref{shape_ngc1667}, showing that this simple model provides a good reproduction of the observations, supporting the scenario of nuclear spiral shocks as drivers of inflows on 100 pc scales.

\abovecaptionskip10ex
\belowcaptionskip2ex

\subsubsection{Connection with AGN feedback}

Inflows towards AGN may also be connected with their outflows, as argued in a recent study of the gas kinematics of the NLR of the Seyfert\,2 galaxy Mrk\,573 \cite{fischer17}, showing that the NLR outflows are related to H$_2$ spiral structures that are likely the source of gas fueling the AGN. Radiation pressure calculations have shown that the strong radiation field of the AGN drives away gas from the proper fueling flow, a mechanism that may apply to outflows seen in other active galaxies. 

Besides feedback due to the AGN radiation pressure, feedback due to accretion disk winds and/or radio jets have also been observed acting on gas that appear to comprise the AGN fuelling flow, including a number of ``equatorial outflows'', observed perpendicularly to the radio jet or ionization axis \cite{riffel14,lena15,sm14a,couto16,couto17}.



\begin{table}
\caption{\bf Masses and mass inflow rates}
\tabcolsep=0.1cm
\begin{tabular}{|l|c|c||}
\hline
                         & Gas mass ($\approx$1\,kpc)$^*$ & Mass inflow rate$^*$ \\
\hline    
\hline    
Cold ($<$100\,K) molecular gas    &$\sim10^6$--10$^8$\,M$_\odot$ &$0.01-10$\,M$_\odot$\,yr$^{-1}$\\
\hline
Warm ($\approx$\,2000\,K) molecular gas  &$\sim50-3000$\,M$_\odot$ &$\sim\,10^{-5}-10^{-3}$\,M$_\odot$\,yr$^{-1}$\\ 
\hline
Hot ($\approx$\,10 000\,K)  ionized gas    &$0.03-8\times10^6$\,M$_\odot$ & $0.01-3\,$M$_\odot$\,yr$^{-1}$\\
\hline
\end{tabular}
\begin{tablenotes}
\item \small $^*$ Uncertainties can be as high as 5--10 times the listed values.
\end{tablenotes}
\end{table} 

\section{Gas masses and mass inflow rates in the near Universe}

We discuss below the values of masses and mass inflow rates obtained in the studies described above. A summarized compilation of the obtained values is presented in Table\,1, where uncertainties of up to factors of 5--10 may apply due to typical uncertainties on the adopted physical and geometric parameters of the flows as well as due to uncertainties in flux calibrations of the IFS data used in the measurements.

Typical values of cold molecular gas masses in AGN hosts, as determined from CO observations of the $\approx$ inner kpc of mostly disk galaxies, are in the range $10^6-10^8$\,M$_\odot$ \cite{schinnerer00,combes14,burillo14,casasola15}. Values in this range have also been observed for the LLAMA (Luminous Local AGN with Matched Analogs) sample \cite{rosario18}, that is also dominated by disk galaxies. 

A larger mass of molecular gas in AGN hosts relative to a control sample of non-active galaxies has been estimated for nearby early-type hosts in a study using Spitzer data \cite{martini13,sl07}: dust masses in the range $M_{\rm dust}\approx\,10^5$--$10^{6.5}$\, M$_\odot$ -- implying gas masses in the range $M_{\rm gas}\approx 10^7$--$10^{8.5}$\, M$_\odot$ --  were found in the AGN, while only  20\% of the control sample of non-active galaxies were detected by Spitzer. If this gas mass in the active galaxies has to be depleted in one activity cycle of 10$^7$--10$^8$\,yr \cite{hickox14} -- as the controls have practically no gas, the implied depletion rate should be in the range $\approx\,0.01-10$\,M$_\odot$\,yr$^{-1}$, either due to inflow of this gas to feed the SMBH at the nucleus and/or to the triggering of star formation in the nuclear region.

Warm molecular gas masses as observed with NIFS and SINFONI within the inner kpc of active galaxies, obtained in the papers discussed in Sec.\,\ref{warm_gas} are much smaller, as they are just the ``warm skin" of the cold molecular gas reservoir, and range between $\sim$\,50 and $\sim$\,3000 M$_\odot$ \cite{riffel18}. Studies in which both the hot and cold molecular gas are observed for the same galaxy \cite{dale05,mazzalay13} give a mean ratio between the masses of cold and hot molecular gas surrounding AGN of  $\approx\,7\times10^5$, although with a large scatter. The AGNIFS group has used this mean scale factor to estimate the cold molecular gas mass expected to be present in the vicinity of the AGN  for which estimates of the warm molecular gas mass are available. Combining these estimated values with other estimates of total molecular gas mass within the inner kpc of AGN hosts result in values between $\approx10^6$\,M$_\odot$ and $10^9$\,M$_\odot$, in approximate agreement with the values obtained with radio telescopes and interferometers such as ALMA. 

The AGNIFS group has also been obtaining masses for the ionized gas within the inner 100--500\,pc, which are in the range (0.03 -- 8)$\times 10^6$\,M$_\odot$ \cite{riffel18} for low-luminosity AGN, while in more luminous ones -- now within the inner kpc -- these masses reach values of 10$^8 - 10^9$\,M$_\odot$ \cite{tadhunter14, couto17}.

The detailed studies of the inner kpc of nearby active galaxies described in the previous sections have led to estimates of mass inflow rates for a few galaxies, although they cannot yet be considered statistically significant results. The difficulty in the quantification of inflows stems from the fact that their kinematic signatures are usually observed as ``residuals" relative to other dominant kinematics, such as rotation, and, in addition, are usually outshined by outflows that are preponderant in the inner few 100 pc of AGN hosts. With these caveats, we have collected below a few numbers from the individual studies discussed above.

The estimated mass inflow rate in NGC\,1566 inferred from the CO gas kinematics observed with ALMA \cite{combes14} is of the order of 10\,M$_\odot$\,yr$^{-1}$. Using a combination of warm molecular gas observations and modelling for the gas mass inflow within the inner few hundred pc of NGC\,1097 \cite{davies09} a mass inflow rate of only 0.05\,M$_\odot$\,yr$^{-1}$ was obtained, while in the case of the more luminous Seyfert 2 galaxy NGC\,1068 \cite{mueller-sanchez09} a mass inflow rate of $\sim$\,15\,M$_\odot$\,yr$^{−1}$ was estimated. Warm molecular gas mass inflow rates, as determined by the AGNIFS group from the observed flux of the H$_2$ molecular line are very small, in the range 10$^{-5}$--10$^{-3}$\,M$_\odot$\,yr$^{-1}$. Nevertheless, as pointed out above, the hot molecular gas is just the warm skin of a larger reservatory of cold molecular gas, thus the actual mass inflow rate in cold gas can be several orders of magnitude higher, most probably reaching a few M$_\odot$\,yr$^{-1}$. Mass inflow rates obtained for the ionized gas in the inflows discussed in Sec.\,\ref{ionized} range from $\approx$0.01 to $\approx$ 3\,M$_\odot$\ yr$^{-1}$. 

Considering that the inflows should be dominated by cold molecular gas, we conclude that the observations of nearby active galaxies support mass inflow rates in the range $\approx 0.01-10$\,M$_\odot$\,yr$^{-1}$, and point out that these values are also in agreement with the values predicted (or fine-tuned to reproduce AGN luminosity functions)  by models \cite{hopkins10,kim17} discussed in \ref{box1} for the inner $\approx$\,300\,pc of gas-rich disk galaxies (similar in particular to that of a model with a bulge-to-disk ratio of B/T=0.3 \cite{hopkins10}).

As a final consideration, we point out that a typical mass-inflow rate of 1\,M$_\odot$\,yr$^{-1}$ is 3 orders of magnitude higher than the typical accretion rates of $\sim10^{-3}$ M$_\odot$\,yr$^{-1}$ of nearby low-luminosity active galaxies.  If this inflow were to last for $10^7$--$10^8$\,yr, gas masses of up to 10$^7$--10$^8$ M$_\odot$ could accumulate in the circumnuclear region, enough to allow the formation new stars. Massive gas reservoirs have been indeed found in the vicinity of AGN, as discussed above, as well recently formed stars, as indeed observed in the inner few 100 pc of many AGN \cite{sb00,raimann05,davies04a,davies06,friedrich10,sb12}. In particular, higher star-formation rates have been found in high-luminosity AGN when compared with control samples \cite{kauffmann03,rosario18,mallmann18}. Another possibility is that at least part of the inflowing gas to the AGN will be ejected via its feedback processes that may act upon the fuelling flow, in the form of strong radiation pressure, accretion disk outflows and radio jets.

We also point out that 3 orders of magnitude is of the order of the ratio between the mass of the bulge and the SMBH mass implied by the M$_{SMBH}$\,vs.\,$\sigma_*$ relation, even though many of the host galaxies are of ``late-type" and seem to have a pseudo-bulge. In any case, these numbers point to some co-evolution of the galaxy and its SMBH.

\section{Summary and Outlook}
\label{final_remarks}

The feeding of SMBHs is a fundamental process leading not only to the growth of the SMBH proper, but intimately related to the evolution of the host galaxy. When observed and quantified, the mass inflow rates are usually much higher than needed to fuel an AGN at the nucleus, and enough to allow the formation of new stars in the central region of the galaxy, leading not only to the SMBH growth but also to the growth of the galaxy. On the other hand, part of this accumulated mass may also be expelled from the nuclear region via the onset of the AGN feedback, as mass outflows rates from AGN on 100--1000 pc scales are estimated to be of the order of the mass inflow rates or even higher.  

\begin{figure}
\hspace{2cm}
\includegraphics[scale=0.50]{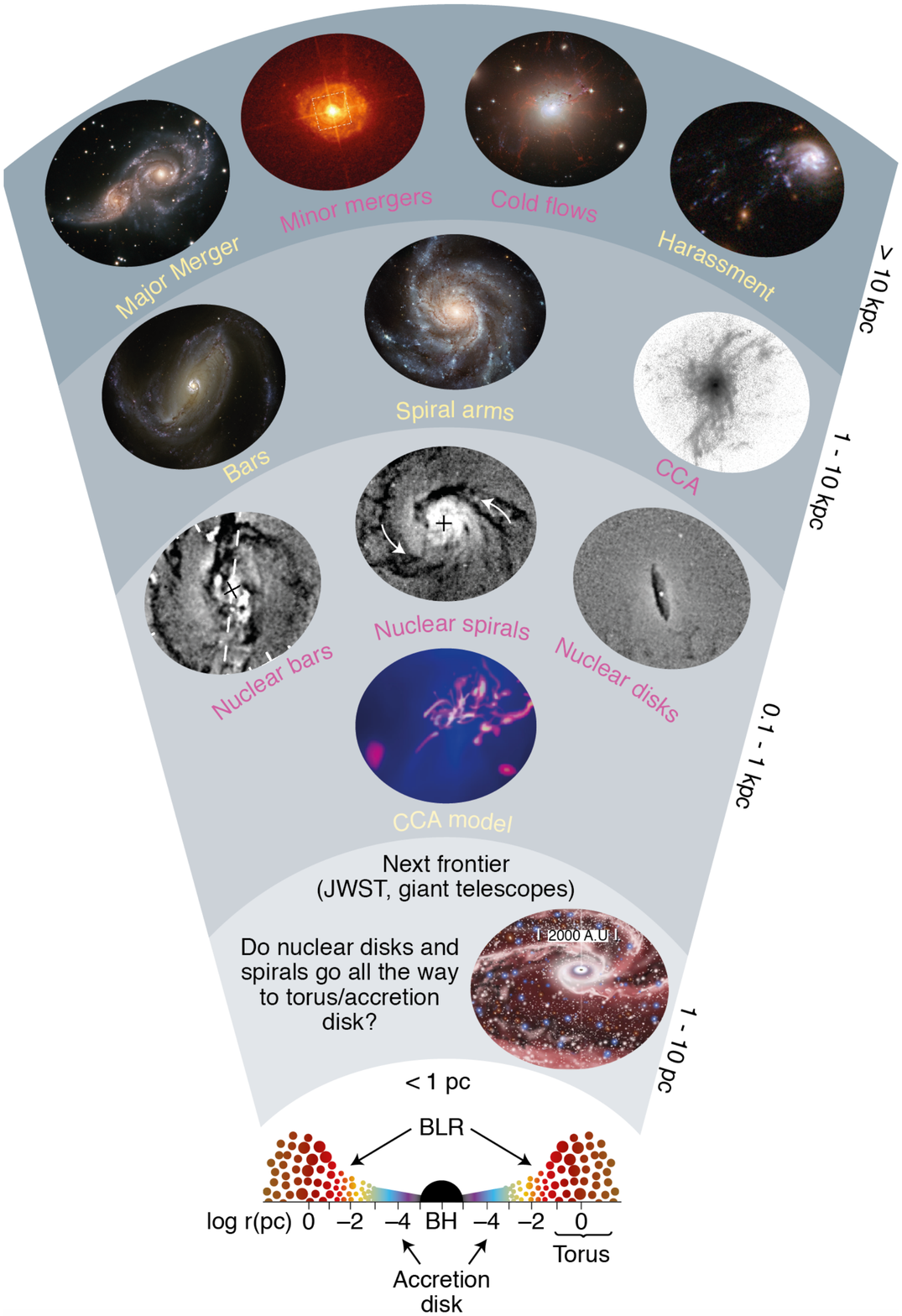}
\caption{{\bf The quest for SMBH feeding:} mechanisms of gas inflow to feed SMBHs at the center of galaxies from extragalactic, through galactic, to 100 pc scales and  the next frontier: resolved observations at parsec scales. CCA means Chaotic Cold Accretion (see text).}
\label{scheme}
\end{figure}

In this review we have discussed many different processes than can feed a SMBH in the center of a galaxy. We have tried to sumarize them in Fig.\,\ref{scheme}, where different tiers show the  processes at play in the feeding of SMBH from the largest to the smallest scales. 

The next frontier of the determination of the mass inflow budget that actually feeds the SMBH lies in the observation of the inner few parsecs, where the SMBH strongly influences the gas -- and stellar -- dynamics (the ``radius of influence" of the SMBH). This region is becoming  observable with new ALMA high resolution observations -- presently reaching 0\farcs015  -- allowing to resolve these inner parsecs in galaxies at distances within $\approx$\,50\,Mpc. The sensitivity of the James Webb Telescope, operating in the infrared, where it can peer inside the obscured central region of AGN hosts, as well as of the giant telescopes of the next decade, operating with adaptative optics, should also provide such observations for the closest active galaxies, including low-luminosity AGN.
\bigskip

\section{Box 1: Theory}
\label{box1}

A thorough theoretical study of AGN fuelling at high luminosities (L$_{AGN}\ge\,10^{46}$erg\,s$^{-1}$) \cite{hopkins10} via hydrodinamical simulations in order to investigate how gas is transported from galactic scales down to accretion disk scales has shown that gas is transported inwards by a series of nested mechanisms, similar to the ``bar within bars'' scenario \cite{schlosman89}. At galactic scales, bars, spiral arms and perturbations induced by galaxy interactions transport gas to the inner $\approx$\,1\,kpc, where gas accumulates. At these scales, once the gas fraction is large enough, the gas disk becomes self-gravitating and vunerable to gravitational instabilities, triggering star formation and gas inflows. These instabilities exhibit diverse morphologies such as nuclear spirals, bars, rings and clumpy discs. They transport gas efficiently to the inner $\approx$\,10\,pc, where another set of instabilities, such as additional spirals, develop and transport gas further inwards. In these simulations, the mass accretion rate is highly variable, reaching $\sim$10\,M$_{\odot}$\,yr$^{-1}$ in an interacting system, and from 0.1 to a few M$_{\odot}$\,yr$^{-1}$ in isolated disc galaxies with a moderate bar instability. Fig.\,\ref{hopkins_10}, shows the gas distribution morphologies resulting from these simulations as a function of bulge and gas masses \cite{hopkins10}. 

More recent hydrodynamic simulations of nuclear gas spirals (within few 100\,pc of the nucleus) in lower luminosity active galaxies \cite{kim17}  (L$_{AGN}\le\,10^{43}$erg\,s$^{-1}$), that are driven by a weak bar-like or oval potential, have shown that  the amplitude of the spirals increase towards the center (inner $\approx$\,80\,pc) where they develop into shocks. These shocks allow gas inflows that can feed a SMBH at accretion rates similar to those observed in nearby Seyfert galaxies. It was found that, if the shear is low, the spirals are losely wound, resulting in a large mass inflow rate, while if the shear is high, the spirals are more tightly wound and form a circumnuclear disk whose gas accretes into the center at a lower rate, as illustrated in Fig.\,\ref{fig:kim}. Typical inflow velocities are of a few tens to 100 km\,s$^{-1}$, in agreement with observations.
\bigskip

\noindent {\it The authors declare no competing financial interests.}\\

\noindent {\it Correspoding author: thaisa@ufrgs.br}

\bigskip
\noindent\textbf{Acknowledgements.}
We acknowledge many colleagues who have contributed with suggestions of discussion points and reference papers which address the feeding of SMBH, specially Richard Davies, Rafaella Morganti, Santiago Garcia-Burillo, Fran\,coise Combes, Witold Maciejewski, Travis Fischer, Michael Crenshaw and David Rosario, besides the anonymous referees who helped to improve the review.

\end{document}